\documentclass[preprint,11pt,nofootinbib]{revtex4}

\usepackage{amsmath}
\usepackage{amsfonts}
\usepackage{amssymb}
\usepackage{colordvi}
\usepackage{float}

\newcommand{\id}{\textrm{d}}
\newcommand{\ran}{\right\rangle}
\newcommand{\lan}{\left\langle}

\newcommand{\be}{\begin{equation}}
\newcommand{\ee}{\end{equation}}
\newcommand{\bea}{\begin{eqnarray}}
\newcommand{\eea}{\end{eqnarray}}

\newcommand{\za}{\alpha}
\newcommand{\zb}{\beta}
\newcommand{\zd}{\delta}
\newcommand{\ze}{\epsilon}
\newcommand{\zg}{\gamma}
\newcommand{\zl}{\lambda}
\newcommand{\zL}{\Lambda}

\newcommand{\zs}{\sigma}
\newcommand{\zt}{\tau}

\newcommand{\zR}{I\hskip-3.4pt R}

\newcommand{\zW}{\Omega}

\newcommand{\zD}{{\Delta}}
\newcommand{\zG}{{\Gamma}}

\newcommand{\WFR}{$\Omega$-FR }

\newcommand{\LFR}{$\Lambda$-FR }
\newcommand{\Wt}{\overline{\Omega}_{t,t+\tau}}
\newcommand{\Ft}{\overline{\mathcal{O}}_{t,t+\tau}}
\newcommand{\Fz}{\overline{\mathcal{O}}_{0,\tau}}

\newcommand{\Wz}{\overline{\Omega}_{0,\tau}}

\newcommand{\noi}{\noindent}

\newcommand {\bdm} {\begin{displaymath}}
\newcommand {\edm} {\end{displaymath}}
\newcommand {\ba}  {\begin{array}}
\newcommand {\ea}  {\end{array}}

\newcommand{\ie}{i.e.\ }

\begin{document}

\title{Focus on some Nonequilibrium Issues}

\author{Matteo Colangeli \email{colangeli@calvino.polito.it}}
\affiliation{Dipartimento di Matematica, Politecnico di Torino, Corso Duca degli Abruzzi
24, I-10129 Torino, Italy}

\author{Lamberto Rondoni}
\affiliation{Dipartimento di Matematica, Politecnico di Torino, Corso Duca degli Abruzzi
24, I-10129 Torino, Italy.\\
INFN, Sezione di Torino, Via P. Giura 1, I-10125, Torino, Italy}

\author{Antonella Verderosa}
\affiliation{Dipartimento di Matematica, Politecnico di Torino, Corso Duca degli Abruzzi
24, I-10129 Torino, Italy}

\begin{abstract}
A mathematical framework for the physics of nonequilibrium phenomena is gradually being developed.
 This review is meant to shed light on some aspects of Response Theory, on the theory of Fluctuation Relations,
 on the so-called \textit{t-mixing} condition, and on the use of Large Deviation techniques in the description of stochastic diffusion processes.
\end{abstract}

\maketitle

\section{Introduction}
Statistical Mechanics provides a mathematical formalism to bridge different scales of investigation of natural phenomena:
a) the microscopic scale, concerning the statistical or collective behaviour of large assemblies of atoms and molecules,
approached e.g.\ in terms of statistical ensembles; b) the mesoscopic
scale, commonly described by the Boltzmann equation and its variations, or by more general and
abstract stochastic processes; c) and the macroscopic level, by and large the realm of Thermodynamics and
Irreversible Thermodynamics which consider matter as a continuum.

Equilibrium phenomena have been investigated and understood much more thoroughly than non-equilibrium
ones. At present, the theory may be considered complete, for what concerns the microscopic foundations
of equilibrium thermodynamics, including the theory of phase transitions and critical phenomena. Differently,
in spite of its celebrated history and of the countless and deep results obtained so far, Statistical Mechanics
has not produced yet a comprehensive theoretical framework for non-equilibrium phenomena. These, indeed, are
much more numerous, diverse and complex than equilibrium phenomena.

Nevertheless, problems posed, in particular, by the modern bio- and nano-technologies, have turned the attention
of a large fraction of the Statistical Mechanics community towards the non-equilibrium phenomena. This has been
possible also thanks to the progress of dynamical systems theory, which becomes necessary when the classical
hypotheses of local equilibrium \cite{mattbook} or kinetic theory fail, as well as in describing macroscopic chaotic phenomena
such as those of turbulence \footnote{Which take place in local equilibrium.}. Indeed, in equilibrium there is no
need to deal with the microscopic dynamics equations
of motion, because the classical ensembles have been proven by experience to accurately capture the statistics
for the macroscopic quantities and their fluctuations. On the contrary, the classical ensembles
do not properly describe systems which are not in equilibrium, in which finite size effects and the persistence of space and time correlations may play a crucial role.
Therefore, new hypotheses and novel approaches are required to describe these systems; in particular, understanding the dynamics of the microscopic
constituents seems to be unavoidable to shed light even on the properties of stationary states.

As a matter of fact, the study of the macroscopic dynamics of dissipative particle systems,
such as those of nonequilibrium molecular
dynamics, has produced a number of results of direct interest in nonequilibrium statistical
mechanics, including relations between transport coefficients and Lyapunov
exponents, which are presently part of a rather satisfactory theory of nonequilibrium liquids. \\
Twenty years ago, the
first fluctuation relation for reversible deterministic dynamics was proposed, and remains one of the few exact
and microscopic results for nonequilibrium systems.
This led to new response formulae,
which generalize the classical response theory to states far from the equilibrium, and to large perturbations
of interest, e.g. in climate studies. Interestingly, various results obtained within the deterministic framework
coincide with those obtained within the stochastic framework, which is reassuring, because in many situations
the two frameworks aim at describing the same phenomenon.

Investigations of Fourier's law of heat conduction have continued along these dynamical lines since the early
days of molecular dynamics and the Fermi-Pasta-Ulam problem, and today they have gained momentum thanks to the
discovery of anomalies in the transport of matter, energy, charge etc. at the nanometric scales, which is of
interest to bio- and nano-technology.

Dynamics and stochastics together may thus advance our understanding of the fundamental principles which are
believed to be common to the incredibly wide spectrum of nonequilibrium phenomena, which
ranges from microscopic to macroscopic scales and includes hydrodynamics and turbulence,
biology, atmospheric physics, granular matter, nanotechnology, etc.

The wealth of techniques developed to tackle the problems of nonequilibrium physics can also be considered as
a theoretical playground for many questions of foundational nature, such as determinism, chaos and randomness,
or emergence and complexity, which find in the problem of irreversibility one of their earliest
examples.

In this paper, we provide a review of some of the cornerstones of nonequilibrium statistical mechanics in order
to clarify the corresponding physical mechanisms. This work is structured as follows.\\
In Sec. \ref{sec1} we analyze the evolution of probability distributions, through the prism of Dynamical Systems theory.\\
In Sec. \ref{sec2}, we address the theory of Linear Response, whose origin can be traced back to the pioneering work of R. Kubo \cite{Kubo}.\\
Section \ref{sec3} focuses on the Onsager-Machlup theory, which concerns the regime of \textit{small} fluctuations around equilibrium.\\
In Sec. \ref{sec4}, we review the theory of Fluctuation Relations.\\
Section \ref{sec5} is devoted to the analysis of the \textit{t-mixing} condition.\\
Section \ref{sec6} presents some results concerning the use of large deviations techniques in stochastic diffusion processes.\\
Conclusions are drawn in Sec. \ref{concl}.

\section{Evolution of probability distributions}
\label{sec1}
This section recalls basic notions of dynamical systems theory, introducing our
notation.
Consider a dynamical system defined by an evolution equation on a phase space $\mathcal{M}$:
\be
\dot{\zG} = F(\zG) ~, \quad \zG \in \mathcal{M}
\label{F}
\ee
whose trajectories for each initial condition $\zG$ are given by $\{ S^t \zG \}_{t\in\mathbb{R}}$,
where $S^t$ is the operator that moves $\zG$ to its position after a time $t$ (e.g. $S^0 \zG=\zG$).
We will consider time reversal invariant dynamics, i.e.\ dynamics obeying
\be
I S^t \zG = S^{-t} I \zG ~, \quad \forall \zG \in \mathcal{M}
\ee
holds, where the linear operator $I : \mathcal{M} \to \mathcal{M}$ is an involution ($I^2\hskip -3pt=$identity)
representing a time reversal operation \footnote{For instance, in simple cases one may take $\zG = ({\bf q},{\bf p})$,
and $I ({\bf q},{\bf p}) = ({\bf q},-{\bf p})$.}. Furthermore, we will consider evolutions such that
$\{ S^t \}_{t\in\mathbb{R}}$ satisfies the group property $S^t S^s = S^{t+s}$.
The time averages of a phase variable $\phi : \mathcal{M} \to \mathbb{R}$, along a trajectory starting at $\zG$,
will be denoted by:
\be
\overline{\phi}(\zG) = \lim_{t \to \infty} \frac{1}{t} \int_0^t \phi\left( S^s \zG \right) {\rm d} s
\ee
If the dynamics represents a thermodynamic system, in which $\zG$ is a single microscopic phase, the time
average should not depend on this phase,
and could be obtained as a phase space average, with respect to a given probability distribution $\mu$
\footnote{Mathematically this condition is verified if the $\Gamma \in \mathcal{M}$ that yield different values
for $\overline{\phi}(\zG)$ constitute a set of vanishing probability. This is a sufficient, not necessary, condition.}:
\be
\overline{\phi}(\zG) = \int_\mathcal{M} \phi(X) ~ {\rm d} \mu(X) = \langle \phi \rangle_\mu ~,
\quad \mbox{for $\mu$-almost every } \zG \in \mathcal{M}
\ee
This is the case if the dynamical system $(S,\mathcal{M},\mu)$ is ergodic (cf.\ Subsection \ref{ErgMix}),
which is a very strong
property, not verified by most of the systems of physical interest. It can be however safely assumed
to hold very often, because physics is often concerned with a small set of observables and with systems made
of exceedingly large numbers of particles, c.f.\ \cite{Khinchin}.

Once $\mathcal{M}$ is endowed with a probability distribution $\mu_0$, $\mu_0(\mathcal{M})=1$ and
$\mu_0(E) \ge 0$ for all allowed events $E \subset \mathcal{M}$, the dynamics in
$\mathcal{M}$ may be used to induce an evolution in the space of probabilities. One may assume
that the subsets of the phase space have a certain probability, which they carry along where
the dynamics moves them. As a consequence, the probability distribution on $\mathcal{M}$ changes
in time, and one may introduce a set of distributions $\{ \mu_t \}_{t \in \mathbb{R}}$ as follows:
\be
\mu_t(E) = \int_E {\rm d} \mu_t = \int_{S^{-t} E} {\rm d} \mu_0 = \mu_0(S^{-t} E)
\label{mutmu0}
\ee
where $S^{-t} E$ is the preimage of $E$ an earlier time $t$. This equation simply means that the
probability of $S^{-t} E$ at the initial time, is assumed to pertain to $E$ at time $t$. With this definition,
probability is conserved in phase space and in general\footnote{In
case of Hamiltonian dynamics, probabilities flow like incompressible fluids.} it flows like a compressible fluid. Taking much care, the evolution
of the probability distributions may be used to define an evolution of the observables, introducing
\be
\langle \phi \rangle_t = \int_\mathcal{M} \phi ~ {\rm d} \mu_t
\ee
Under certain conditions, the mean values of the phase functions completely characterize the system, therefore one often refers to
$\mu_t$ as to the {\em state} of the system at time $t$, which is to be distinguished from the microscopical phase
$\Gamma \in \mathcal{M}$. A probability measure $\mu$ is called
{\em invariant} if $\mu(E)=\mu(S^{-t} E)$ for all $t$ and all measurable sets $E$.

At times $\mu_t$ has a density $f_t$, i.e.\
${\rm d} \mu_t(\zG) = f_t(\zG) {\rm d} \zG$. In that case, the evolution of $\mu_t$
follows from the evolution of the normalized non-negative function $f_t$, determined by
Eq.(\ref{mutmu0}). Operating in Eq.(\ref{mutmu0}) the change of coordinates $Y = S^t X$, i.e.\ $X=S^{-t} Y$, in the last integral
of the following expression
\be
\mu_t(E) = \int_E f_t(X) ~ {\rm d} X = \int_{S^{-t} E} f_0(X) ~ {\rm d} X
\ee
and:
\be
\int_E f_t(X) ~ {\rm d} X = \int_{E} f_0(Y) J^{-t}(Y) ~ {\rm d} Y
\label{EftX}
\ee
where $J^{-t}(Y) = |(\partial S^{-t} X / \partial X)|_Y$ is the Jacobian of the transformation.
As Eqs.(\ref{mutmu0}-\ref{EftX}) hold for all allowed subsets of $\mathcal{M}$, one can write
\be
f_t(X) = f_0(S^{-t}X) J^{-t}(X)
\label{fEvol}
\ee
For Hamiltonian dynamics, $J^{-t}(X)=1$, hence $f_t(X) = f_0(S^{-t}X)$.
In general, for the evolution of the observables one obtains:
\be
\langle \phi \rangle_t = \int_\mathcal{M} \phi(\zG) f_t(\zG) {\rm d} \zG =
\int_\mathcal{M} \phi(\zG) f_0(S^{-t} \zG) J^{-t}(\zG) {\rm d} \zG
\ee
Introducing $Y=S^{-t} \zG$ in the last integral, so that ${\rm d} \zG = J^t(Y) {\rm d} Y$, one finds:
\be
\langle \phi \rangle_t = \int_\mathcal{M} \phi(S^t Y) f_0(Y) J^{-t}(S^tY) J^t(Y) {\rm d} Y
\label{Aevolv}
\ee
Under suitable smoothness conditions for the dynamics and $\mathcal{M}$, probability is transported by the phase space points
like the mass of a fluid, whose density $f$ obeys the formal
continuity equation:
\be
\frac{\partial f}{\partial t} = - \nabla_\zG \cdot \left(F f \right)~, \quad
\frac{{\rm d} f}{{\rm d} t} = \frac{\partial f}{\partial t} + \nabla_\zG f \cdot F= - f \nabla_\zG \cdot F = - f \zL
\label{Liouville}
\ee
Here $\zL=\nabla_\zG \cdot F$, called phase space expansion rate,
is the divergence of the vector field $F$ on $\mathcal{M}$, cf.\ Eq.(\ref{F}).
Introducing the total time derivative
\be
\frac{\rm d}{{\rm d} t} = \frac{\partial}{\partial t} + F \cdot \nabla_\zG ~,
\ee
Eqs.(\ref{Liouville}) may also be written as
\be
\frac{\rm d}{{\rm d} t} \ln f = -\zL
\label{totde}
\ee
Because the global existence and uniqueness of solutions of the equations of motion is practically
assured for particle systems of physical interest,\footnote{Global solution means that particles do no
cease to exist after a while; Uniqueness implies that the same particles do not exist at once
along distinct trajectories. If these properties are violated, the model under investigation must be discarded.}
 one may safely assume that the solutions
of the Liouville equation also exist and can be constructed by means of formal calculations.
Various procedures are available for this purpose. For example, let us introduce the $f$-Liouvillean
operator ${\cal L}$:
\be
{\cal L} = -i \left( \nabla_\zG \cdot F + F \cdot \nabla_\zG \right)~, \quad \mbox{so that } ~~
\frac{\partial f}{\partial t} = - i {\cal L} f
\ee
and let us express $\partial f_t / \partial t$ to first order in the time increment $\Delta t$:
\be
\frac{\partial f_t}{\partial t}(\zG) = -i \left( {\cal L} f_t \right) (\zG) =
\frac{f_{t+\zD t}(\zG) - f_t(\zG)}{\zD t} + O\left(\zD t \right)
\ee
It follows that
\bea
&&f_{\zD t}(\zG) = \left( 1 - i {\cal L} \zD t \right) f_0 (\zG) + O\left(\zD t^2 \right) \\
&&f_{2\zD t}(\zG) = \left( 1 - i {\cal L} \zD t \right) f_{\zD t} (\zG) + O\left(\zD t^2 \right) =
\left( 1 - i {\cal L} \zD t \right)^2 f_0 (\zD) + O\left(\zD t^2 \right)  \\
&& \hskip 37pt {\vdots} \\
&&f_{n \zD t}(\zG) = \left( 1 - i {\cal L} \zD t \right)^n f_0 (\zG) + n O\left(\zD t^2 \right)
\eea
Taking $\zD t=t/n$, so that $\zD \to 0$ and $n O\left(\zD t^2 \right) \to 0$ as $n \to \infty$,
one obtains:
\be
f_t(\zG) = \lim_{n \to \infty} \left( 1 - \frac{i t {\cal L}}{n} \right)^n f_0(\zG) =
\sum_{n=0}^\infty \frac{\left( -i t {\cal L} \right)^n}{n!} f_0(\zG) \equiv e^{-i t {\cal L}} f_0(\zG)
\label{fLiouv}
\ee
The question is now to connect Eq.(\ref{fLiouv}) with Eq.(\ref{fEvol}).
One can write
\be
Y = S^t X = S^{t/n} \left( S^{t/n} \left( \cdots S^{t/n} \left( X \right) \cdots  \right) \right)
\ee
Hence, the chain rule yields
\be
\left. \frac{\partial Y}{\partial X}\right|_{X_i} =
\left( \left. \frac{\partial S^{t/n} X}{\partial X}\right|_{X_{n-1}} \right)
\left( \left. \frac{\partial S^{t/n} X}{\partial X}\right|_{X_{n-2}} \right) \cdots
\left( \left. \frac{\partial S^{t/n} X}{\partial X}\right|_{X_0} \right)
\label{chainr}
\ee
where $X_{j}=S^{j t/n} X_{0}$, and $X_{0}$ is the initial point of a trajectory.
One can expand to first order each derivative in brackets as follows:
\be
\left. \frac{\partial \left(S^{t/n} X\right)}{\partial X}\right|_{X_j}
= \left. \frac{\partial}{\partial X}\left(X + F \zD t + O(\zD t^2))\right)\right|_{X_j}
\ee
and further
\bea
\left. \frac{\partial \left(S^{t/n} X\right)}{\partial X}\right|_{X_j}= 1 + \left. \frac{\partial F}{\partial X}\right|_{X_j} \zD t + O\left( \zD t^2 \right)
= e^{\left. \frac{\partial F}{\partial X}\right|_{X_j} \zD t} + O\left( \zD t^2 \right)~,
\label{pst}
\eea
$1$ being the identity matrix.
Substituting Eq.(\ref{pst}) in Eq.(\ref{chainr}),
and noting that the exponential operators do not commute in general, the $n \to \infty$ limit
leads to a so-called \emph{left ordered} exponential, which can also be expressed as a Dyson series:
\bea
e_L^{\int_0^t T(S^s X) {\rm d} s} &=& 1 + \int_0^t {\rm d} t_1 ~ T(S^{t_1} X) +
\int_0^t {\rm d} t_1 \int_0^{t_1} {\rm d} t_2 ~ T(S^{t_1} X) T(S^{t_2} X)  \nonumber \\
&&+\int_0^t {\rm d} t_1 \int_0^{t_1} {\rm d} t_2 \int_0^{t_2} {\rm d} t_3 ~
T(S^{t_1} X) T(S^{t_2} X) T(S^{t_3} X) + \dots  \nonumber
 \eea
where the time dependent matrix
\be
T(S^s X)= \left. \frac{\partial F}{\partial X} \right|_{S^sX}
\ee
is the Jacobian matrix of $F$ computed at the point $S^s X$. Considering that the identity
det$(e^L)=\exp($Tr$L)$ holds for left ordered exponentials as well, one obtains:
\be
\det \left( e_L^{\int_0^t T(S^s X) {\rm d} s} \right) = \exp \left\{
\int_0^t \nabla_\zG \cdot F \left( S^s X \right) {\rm d} s \right\} =
\int_0^t \zL \left( S^s X \right) {\rm d} s
\ee
Which implies that:
\be
J^t(X) = e^{\int_0^t \zL(S^u X) {\rm d} u} =
e^{\int_{-t}^0 \zL(S^{t+s} X) {\rm d} s} = \frac{1}{J^{-t}\left( S^t X \right)} =
 \frac{1}{J^{-t}\left( Y \right)}
\label{jacob}
\ee
where we have taken $u=t+s$ in the second integral. Equation (\ref{jacob}) is obvious for
compressible fluids: a fluid element about $X$ varies in a time $t$ by a factor which
is the inverse of the variation of the fluid element about $Y$, when tracing backwards its
trajectory. Consequently $J^{-t}(S^t X) J^t(X)=1$, and Eq.(\ref{fEvol})
may be rewritten as:
\be
f_t(X) = f_0(S^{-t}X) e^{-\int_{-t}^0 \zL(S^s X) {\rm d} s}
\label{Evolf}
\ee
while Eq.(\ref{Aevolv}) takes the interesting form
\be
\langle \phi \rangle_t = \int_\mathcal{M} \left(\phi \circ S^t\right)\hskip -2pt(X)~
f_0(X) ~ {\rm d} X
=\langle \phi \circ S^t \rangle_0
\label{useful}
\ee

\subsection{Ergodicity and mixing}
\label{ErgMix}
Let $\mu$ be one invariant probability
 distribution and $\phi$ an integrable phase function. The following statements are equivalent:
\begin{itemize}
\item[\bf E1.] $\overline{\phi}(\zG) = \langle \phi \rangle_\mu$,
except for a set of vanishing $\mu$ probability;
\item[\bf E2.] except for a set of vanishing $\mu$ probability,
$\tau_E(\zG)=\mu(E)$, where $E \subset \mathcal{M}$ is a $\mu$-measurable set and
\be
\tau_E(\zG) = \lim_{t \to \infty} \frac{1}{t} \int_0^t \chi_E \left( S^s \zG \right) {\rm d} s ~;
\quad \mbox{with } ~\chi_E \left( \zG \right) =
\left\{ \begin{array}{ll}
1 ~ & \mbox{if } \zG \in E \\
0 ~ &  \mbox{else } \end{array} \right.
\ee
is the the mean time in $E$;
\item[\bf E3.] let $\phi$ be $\mu$-integrable and let $\phi$ be a constant of motion
(i.e.$\phi(S^t \zG) = \phi(\zG)$ for all $t$ and
all $\zG$). Then $\phi(\zG) = C$ $\mu$-almost everywhere, for a given $C \in \mathbb{R}$;
\item[\bf E4.] the dynamical system $(S,\mathcal{M},\mu)$ is {\em metrically indecomposable}, i.e.\
given the invariant set $E$ (which means $S^{-t} E = E$), either $\mu(E)=0$ or $\mu(E)=1$.
\end{itemize}
We call {\em ergodic} the dynamical systems that verify these statements. This is a very strong property because $\phi$ can be any integrable function.
Physics concerns, instead, only a few phase variables that are physically relevant. \\
The following statements are equivalent too:
\begin{itemize}
\item[\bf M1.] For every pair of measurable sets $D,E \subset \mathcal{M}$ one has:
\be
\lim_{t \to \infty} \mu \left( S^{-t} D \cap E \right) = \mu(D) \mu(E)
\ee
\item[\bf M2.] for all $\phi , \psi \in L_2 (\mathcal{M},\mu )$ the following holds:
\be
\lim_{t \to \infty} \left\langle \left( \phi \circ S^t \right) \psi \right\rangle_\mu =
\left\langle \phi \right\rangle_\mu \left\langle \psi \right\rangle_\mu
\ee
\end{itemize}
We call {\em mixing} the dynamical systems that verify these two statements. Mixing is an even stronger
property than ergodicity, in the sense that mixing systems are also ergodic, whereas not all
ergodic systems are mixing.

For dynamics, which are mixing with respect to a probability measure with density $h$,
$d \mu = h d\Gamma$ say,  one can prove that an initial state characterized by a
probability density $f_0$ eventually converges to the state of density $h$. To prove
that, consider the phase functions $\phi$ and $\psi$, for which one can write:
\bea
&&\hskip -30pt \lim_{t\to\infty}
\left\langle (\phi \circ S^t) \cdot \psi \right\rangle_h =
\lim_{t\to\infty}
\left\langle \phi \circ S^t \right\rangle_h \left\langle \psi \right\rangle_h \nonumber \\
&&= \left\langle \psi \right\rangle_h \int d\Gamma \phi(S^t \Gamma)h(\Gamma) =
\left\langle \psi \right\rangle_h \int d\Gamma \phi (\Gamma) {S^{*}}^t h(\Gamma)
=\left\langle \psi \right\rangle_h \left\langle \phi \right\rangle_h \nonumber
\eea
where the superscript $^*$ denotes the distribution function propagator for a period time $t$.
Then, for a time dependent probability distribution $f_t$ which vanishes at least where $h$
does, let us introduce $R_t=f_t / h$:
\be
\hskip -0.2cm \int R_t(\Gamma) h(\Gamma)d\Gamma =\int f_t(\Gamma) d\Gamma=1;
\mbox{ }
\int \frac{1}{R_t(\Gamma)}f_t(\Gamma)d\Gamma=\int h(\Gamma)d\Gamma=1
\ee
for all times $t$, and we obtain:
\be
\left\langle \phi \right\rangle_t = \int \phi (\Gamma)f_t(\Gamma)d \Gamma=
\int \phi (\Gamma)R_t(\Gamma)h(\Gamma)d\Gamma= \left\langle \phi \cdot R_t \right\rangle_h
\ee
We can also write, by definition:
\be
\left\langle \phi \right\rangle_t = \int \phi (\Gamma)f_t(\Gamma)d \Gamma=
\int \phi (S^t \Gamma) f_0(\Gamma)d \Gamma=\int \phi(S^t \Gamma)R_0(\Gamma)h(\Gamma)d \Gamma
\ee
from which, the mixing condition produces the convergence to the steady state of density $h$:
\be
\lim_{t \to \infty} \left\langle \phi \right\rangle_t=\int \phi(S^t \Gamma) R_0(\Gamma)h(\Gamma)d \Gamma=
\left\langle \left( \phi \circ S^t \right) R_0 \right\rangle_h \rightarrow
\left\langle \phi \right\rangle_h  \left\langle R_0 \right\rangle_h= \left\langle \phi \right\rangle_h
\ee
In other words, probability densities for finite systems, if they are both stationary
and mixing, are attractors in the space of probability densities.

However, this proof of convergence to a mixing stationary state is deceitfully simple.
Although it is a very strong property, in general mixing does not
suffice to prove convergence to a steady state, because it amounts to the decay in time of the microscopic
correlations within already stationary macroscopic states and not to the decorrelation of the initial state
from the final state.

\section{Linear Response}
\label{sec2}
Let us address the response of a given system
to external actions. As an example, consider
a system of $N$ particles in contact with a thermal bath at inverse temperature
$\zb$, described by the following Hamiltonian:
\be
H(\zG) = H_0(\zG) + \zl A(\zG)~,
\ee
where $\zl$ is a small parameter and $A$ perturbs the canonical equilibrium:
\be
f_0 = \exp (-\zb H_0) / \int {\rm d} \zG \exp (-\zb H_0)
\ee
After some time, a new canonical
equilibrium is established which, to the first order in $\zl$, is given by:
\bea
f = \frac{e^{-\zb H_0} e^{-\zb \zl A}}{\int {\rm d} \zG e^{-\zb H_0} e^{-\zb \zl A}}
= \frac{e^{-\zb H_0} \left[ 1 - \zb \zl A + O(\zb^2 \zl^2 A^2) \right]}{\int {\rm d} \zG e^{-\zb H_0} \left[1 - \zb \zl A + O(\zb^2 \zl^2 A^2) \right]}
\nonumber \\
\simeq f_0 \frac{1-\zl \zb A}{1-\zl \zb \left\langle A \right\rangle_0}
\simeq f_0(\zG) \left[ 1 - \zl \zb \left( A(\zG) - \left\langle A \right\rangle_0 \right) \right]
\eea
where, $\langle \cdot \rangle_0$ denotes averaging with respect to $f_0$. The effect of the perturbation
on a given observable $\phi$, is then expressed by:
\be
\langle \phi \rangle_{\lambda} - \langle \phi \rangle_0 = \int {\rm d} \zG \phi(\zG) \left[ f(\zG) - f_0(\zG) \right]
\simeq -\zl \zb \left[ \langle \phi A \rangle_0 - \langle \phi \rangle_0 \langle A \rangle_0  \right]
\label{perturbed.avdiff}
\ee
which is the correlation of the observable $\phi$ with the perturbation $A$, with respect to the state
expressed by $f_0$.
Taking $\phi=A=H_0$, one obtains an expression for the heat capacity at constant
volume $C_V$, which expresses the response of the system to temperature variations.
Indeed, defining $C_V$ as
\be
C_V = \frac{\partial \langle H_0 \rangle_0}{\partial T}=
\frac{{\rm d} \zb}{{\rm d}T} \frac{\partial \langle H_0 \rangle_0}{\partial \zb} =
\frac{\langle H_0^2 \rangle_0 - \langle H_0 \rangle_0^2}{k_{_B} T^2}
\label{cv}
\ee
Eqs (\ref{perturbed.avdiff},\ref{cv}) yield:
\be
\left. \frac{\partial \langle H_0 \rangle}{\partial \zl} \right|_{\zl=0} = \lim_{\zl \to 0}
\frac{\langle H_0 \rangle_{\zl} - \langle H_0 \rangle_0}{\zl} =
- \zb \left[ \langle H_0^2 \rangle_0 - \langle H_0 \rangle_0^2  \right] = - k_{_B} T^2 C_V
\ee
More in general, consider time dependent perturbations of form $-{\cal F}(t) A(\zG)$:
\be
H(\zG,t) = H_0(\zG) - {\cal F}(t) A(\zG)
\ee
and split the corresponding evolution operator in two parts:
\be
i {\cal L}_0 f = \left\{ f,H_0 \right\} ~, \quad i {\cal L}_{\rm ext}(t) f =
- {\cal F}(t) \left\{ f, A \right\}
\ee
where $\{ \cdot \}$ are the Poisson brackets. One has $i {\cal L}_0 f_0 = 0$, which means
that $f_0$ is invariant for the unperturbed dynamics. Then, the solution of the Liouville
equation
\be
\frac{\partial f}{\partial t} = -i \left( {\cal L}_0 + {\cal L}_{\rm ext}(t) \right) f
\ee
can be expressed by:
\bea
f_t(\zG) &=& e^{i t {\cal L}_0} f_0(\zG)  - i \int_0^t {\rm d} t' e^{-i(t-t'){\cal L}_0}
{\cal L}_{\rm ext}(t') f_{t'}(\zG) \nonumber \\
&=& f_0(\zG)  - i \int_0^t {\rm d} t' e^{-i(t-t'){\cal L}_0} {\cal L}_{\rm ext}(t') f_0(\zG)
+ \mbox{higher order in}~{\cal {\cal L}_{\rm ext}} \nonumber
\eea
as proved by inspection.
If the deviations from the unperturbed system are considered small, the higher orders in ${\cal L}_{\rm ext}$ can be omitted. Then Eq.(\ref{perturbed.avdiff}) implies:
\be
\langle \phi \rangle_t - \langle \phi \rangle_0 \simeq \int {\rm d} \zG \phi(\zG)
\int_0^t {\rm d} t' e^{-i(t-t'){\cal L}_0} {\cal F}(t') \left\{ f_0 , A \right\}
\ee
where
\be
\left\{ f_0 , A \right\}=\left\{ H_0 , A \right\} \frac{\partial f_0}{\partial H_0} =
\zb f_0 \frac{{\rm d} A}{{\rm d} t} \label{rham}
\ee
Eventually, one obtains:
\be
\langle \phi \rangle_t - \langle \phi \rangle_0 \simeq \int_0^t {\rm d} t' R(t-t') {\cal F}(t')
\label{trespo}
\ee
where $R(t)$ is the response function:
\be
R(t) = \zb \left\langle \dot{A} \left( \phi \circ S^t \right) \right\rangle_0 =
\zb \int {\rm d} \zG f_0(\zG)  \frac{{\rm d} A}{{\rm d} t}(\zG) e^{i t {\cal F}_0} \phi(\zG)
\ee
Once again, the macroscopic nonequilibrium behaviour of a given system has been related
solely to the correlations of microscopic fluctuating quantities, computed with respect to
the relevant equilibrium ensemble.

Equation (\ref{trespo}) suggests that even the linear response is in general affected by
memory effects, hence the Markovian behaviour appears to be either very special or only
approximately valid.  This implies, for instance, that all nonequilibrium fluids have a
viscoelastic behaviour.  In practice, however,  in normal fluids this  behaviour arises
only exceedingly far from equilibrium.

Recently, it has been shown that
this approach applies to the case of perturbation of non equilibrium steady states, if they are represented by a regular
probability density, as in the presence of noise, cf.\ Refs.\cite{BPRV,FIVb}.

Differently, the invariant phase space probability distribution $\mu$ of a dissipative system
 is singular and supported on a fractal attractor. Consequently, it
is not obvious anymore that the statistical features induced by a perturbation can be
related to the unperturbed statistics. The reason is that even very small perturbations may
lead to microscopic phase whose probability vanishes in the unperturbed state.
In such a case, the information contained in $\mu$ is irrelevant.

Indeed, Ruelle \cite{Ruelle} showed that in certain cases\footnote{Concerning certain smooth, uniformly
hyperbolic dynamical systems.} a perturbation $\delta \zG$ about
a microstate $\zG$ and its evolution $S^t \delta \zG$ can be decomposed in two
parts, $(S^t \delta \zG)_{\parallel}$ and $(S^t \delta \zG)_{\perp}$, respectively
perpendicular and parallel to the fibres of the attractor:
$$
S^t \delta \zG = (S^t \delta \zG)_{\parallel}+(S^t \delta \zG)_{\perp}
$$
The first addend can be related to the dynamics on the attractor, while the second
may not.

Later, it has been pointed out \cite{CRV12} that this difficulty should not concern
systems of many interacting particles.
In those cases, rather than
the full phase space, one considers the much lower dimensional projections concerning the few
physically relevant observables, i.e. the marginals of singular
phase space measures, on spaces of sufficiently lower dimension, which are usually regular \cite{EvRon,BKL2005}.
These facts can be briefly recalled as follows.
Ruelle showed that the effect of a perturbation
$\delta F(t)=\delta F_{\parallel}(t)+\delta F_{\perp}(t)$ on the response of a generic
(smooth enough) observable $\phi$ is given by:
\be
\langle{\phi}\rangle_t - \langle{\phi}\rangle_0 = \int_{0}^{t} R_{\parallel}^{(\phi)}(t-\tau)
\delta F_{\parallel}(\tau) d\tau+\int_{0}^{t} R_{\perp}^{(\phi)}(t-\tau) \delta F_{\perp}(\tau) d\tau
\label{4}
\ee
where the subscript 0 denotes averaging with respect to $\mu$,
$R_{\parallel}^{(\phi)}$ may be expressed in terms of
correlation functions evaluated with respect to $\mu$, while
$R_{\perp}^{(\phi)}$ depends on the dynamics along the stable manifold, hence it may not.

Let us adopt the point of view of Ref.\cite{CRV12}. For a $d$-dimensional
dissipative dynamical system consider, for simplicity, an impulsive perturbation
$\zG \to \zG + \zd \zG$, such that all components of $\zd \zG$ vanish except one, denoted by
$\zd \zG_i$. The probability distribution $\mu$ is correspondingly shifted by $\zd \zG$, and
turns into a non-invariant distribution $\mu_0$, whose evolution $\mu_t$ tends to $\mu$ in the
$t \to \infty$ limit. For every measurable set
$E \subset {\cal M}$, $\mu_0(E)$ equals $\mu(E - \zd \zG)$,\footnote{The set
$E - \zd \zG$ is defined by $\{ \zG \in {\cal M} : \zG + \zd \zG \in E \}$.}
and $\mu_t(E)$ is computed as explained in Sec.~\ref{sec1}. Taking
$\phi(\zG) = \zG_i$, one obtains:
\be
\langle{\zG_i}\rangle_t - \langle{\zG_i}\rangle_0 =
\int \zG_i ~ {\rm d} \mu_t(\zG) - \int \zG_i ~ {\rm d} \mu(\zG)
\label{meas1}
\ee
Let us now approximate the singular $\mu$, coarse graining $\mathcal M$ with an $\epsilon$-partition made of
a finite set of $d$-dimensional hypercubes $\Lambda_k(\epsilon)$ of side $\epsilon$ and
centers $\zG_k$. The corresponding approximations of $\mu$ and of $\mu_t$ are given by
 the probabilities $P_k(\epsilon)$ and $P_{t,k}(\epsilon;\delta \zG)$
of the hypercubes $\Lambda_k(\epsilon)$, where:
\be
P_k(\epsilon)=\int_{\Lambda_k(\epsilon)} {\rm d} \mu(\zG) \hskip 3pt, \quad
P_{t,k}(\epsilon)
= \int_{\Lambda_k(\epsilon)} {\rm d} \mu_t(\zG) ~.
\label{CG}
\ee
The coarse grained invariant density $\rho(\zG;\epsilon)$ is given by:
\be
\rho(\zG;\epsilon) =
\sum_k \rho_k(\zG;\epsilon) \hskip 3pt, \hskip 6pt \hbox{with} \hskip 6pt
\rho_k(\zG;\epsilon) = \left\{
\begin{array}{ll}
    P_k(\epsilon)/\epsilon^d & \hbox{if $ x \in \Lambda_k(\epsilon)$} \\
    0 & \hbox{else}\label{densCG}
  \end{array}\right.
\ee
If $Z_i$ is the number of one-dmensional bins of form
$\left[\zG_i^{(q)}-\epsilon/2,\zG_i^{(q)}+\epsilon/2\right)$, $q\in\{1,2,...,Z_i\}$,
in the $i$-th direction, marginalizing the approximate distribution
yields the quantities:
\be
p_i^{(q)}(\epsilon) =
\int_{\zG_i^{(q)}-\frac{\epsilon}{2}}^{\zG_i^{(q)}+\frac{\epsilon}{2}}
\left\{\int \rho(\zG;\epsilon) \prod_{j\neq i} {\rm d} \zG_j \right\}
{\rm d} \zG_i
\label{CG2} ,
\ee
each of which is the invariant probability that the coordinate $\zG_i$ of $\zG$
lie in one of the $Z_i$ bins.
Similarly, one gets the marginal of the evolving approximate probability:
\be
p_{i,t}^{(q)}(\epsilon) =
\int_{\zG_i^{(q)}-\frac{\epsilon}{2}}^{\zG_i^{(q)}+\frac{\epsilon}{2}}
\left\{\int \rho_t(\zG;\epsilon) \prod_{j\neq i} {\rm d} \zG_j \right\}
{\rm d} \zG_i
\label{CG3} ,
\ee
Dividing by $\ze$, one obtains the coarse grained marginal probability densities
$\rho_i^{(q)}(\epsilon)$ and $\rho_{t,i}^{(q)}(\epsilon)$, as well as the
$\epsilon$-approximate response function:
\be
B_{i}^{(q)}(\zG_i,\delta \zG,t,\epsilon)
= \frac{1}{\epsilon} \left[p_{t,i}^{(q)}(\epsilon)-p_i^{(q)}(\epsilon)\right]
= {\rho_{t,i}^{(q)}(\epsilon)-\rho_i^{(q)}(\epsilon)}
\label{CG4}
\ee
Reference \cite{CRV12} shows that the right hand side of Eq.(\ref{CG4}) tends to a
regular function of $\zG_i$ under the $Z_i \rightarrow \infty$, $\epsilon \to 0$
limits. Consequently, $B_{i}^{(q)}(\zG_i,\delta \zG,t,\epsilon)$ yields
an expression similar to that of standard response theory, in the sense that it
depends solely on the unperturbed state, although that is supported on a fractal set. There are
exceptions to this conclusion, most notably those discussed by Ruelle. But for most systems of physical interest,
such as systems of many interacting particles, this is the expected result.
The idea is that the projection procedure makes unnecessary
the explicit calculation of $R_{\perp}^{(\phi)}$ in Eq.(\ref{4}), although
 $R_{\perp}^{(\phi)}$ does not need to be negligible \cite{Sepul}. Therefore, apart from
peculiar situations, the response may
be referred only to the unperturbed dynamics, as in the standard theory.

\section{Onsager-Machlup: response from small deviations}
\label{sec3}
The classical theory of fluctuations, developed by Onsager and Machlup \cite{OM53a,OM53b}
 to quantify the probability of temporal fluctuations paths, is based on the following assumptions:
\begin{itemize}
\item[\bf A1.] Onsager regression hyptothesis: the decay of a system from a nonequilibrium
state produced by a spontaneous fluctuation, obeys on average the macroscopic law
describing the decay from the same state produced by a macroscopic
constraint that has been suddenly removed;
\item[\bf A2.] the observables are Gaussian random variables (i.e.\ the probability
density of $m$ values taken at $m$ consecutive instants of time is an $m$-dimensional
Gaussian);
\item[\bf A3.] the probability density $P(\zG)$ of the microstate $\zG$ obeys Boltzmann's principle:
\be
k_{_B} \log P(\zG) = {\cal S}(\zG) + {\rm const}
\ee
\item[\bf A4.] the state $S^t\zG$ is statistically independent of the state $S^{t'}\zG$
for $|t-t'| > \zt_d$, $\zt_d$ being the decorrelation time;
\item[\bf A5.] the microscopic dynamics is time reversal invariant;
\item[\bf A6.] the vector of observables $\za=(\za_1,...,\za_n)$ is chosen so that its
evolution is Markovian. This is possible if $n$ is neither too small nor too large in such a way that:
\begin{itemize}
\item $\za_i$ represents a macroscopic quantity referring to
 a subsystem containing very many particles;
\item $\za_i$ is an algebraic sum of molecular variables, so that by the Central Limit
Theorem its fluctuations are Gaussians centered on its average
(equilibrium) value;
\item $\za_i$ must be an even function of the molecular variables that are odd under
time reversal (microscopic time reversal invariance);
\end{itemize}
\item[\bf A7.] the system is in local thermodynamic equilibrium;
\item[\bf A8.] the fluxes $\dot{\za}_i$ depend
linearly on the thermodynamic forces $X_i$:
\be
\dot{\za}_i = \sum_{j=1}^n L_{ij} X_j ~, \qquad X_i = \sum_{j=1}^n R_{ij} \dot{\za}_j ~;
\ee
\item[\bf A9.] the process is stationary: i.e.\ given the times $t_1, t_2, ..., t_p$ and the
$n$-dimensional vectors $\za^{(1)}, \za^{(2)}, ..., \za^{(p)}$, the
probabilities $F_{i,p},~ i=1,...,n$, that each component of the observable vector
is smaller by value than the corresponding component of the vector sequence $\za^{(k)}$
at the corresponding times $t_k$ satisfy:
\be
F_{i,p}\left(\za_i \le \za_i^{(k)} , t_k , k=1,...,p\right) =
F_{i,p}\left(\za_i \le \za_i^{(k)} , t_k+\zt, k=1,...,p\right)
\ee
for all $\zt$ and, analogously, the corresponding probability densities $f_{i,p}$, satisfy
\be
f_{i,p}\left(\za_i = \za_i^{(k)} , t_k,  k=1,...,p\right) =
f_{i,p}\left(\za_i = \za_i^{(k)} , t_k+\zt, k=1,...,p\right)
\ee
where
\bea
&&F_{i,p}\left(\za_i \le \za_i^{(k)} , t_k , k=1,...,p\right) \nonumber \\
&& \hskip 20pt =
\int_{-\infty}^{\za_i^{(1)}} {\rm d} \za_i^{(1)} \cdots \int_{-\infty}^{\za_i^{(p)}} {\rm d} \za_i^{(p)}
f_{i,p}\left(\za_i = \za_i^{(k)} , t_k , k=1,...,p\right) \nonumber
\eea
\end{itemize}
For simplicity, let $\za$ be the vector of the deviations from the equilibrium values.
Then, the entropy ${\cal S}$ is a function of the observables $\za$, which can be expanded
about its equilibrium value ${\cal S}_0$ as:
\be
{\cal S} = {\cal S}_0 -\frac 1 2 \sum_{i,j=1}^n s_{ij} \za_i \za_j + \mbox{higher order in } \za
\ee
There is no linear term in $\za$ because ${\cal S}_0$ is the maximum of ${\cal S}$.
Correspondingly, the thermodynamic forces are expressed by
\be
X_i = \frac{\partial {\cal S}}{\partial \za_i} = - \sum_{j=1}^n s_{ij} \za_j ~, \qquad i=1,...,n
\ee
which implies
\be
\sum_{j=1}^n \left[ R_{ij}\dot{\za}_j + s_{ij} \za_j \right] = 0 ~, \qquad i=1,...,n
\label{hydrod}
\ee
To compute the evolution of $\za$, let us introduce the functions
\be
\Phi\left( \dot \za , \dot \zb \right) = \frac 1 2 \sum_{i,j=1}^n R_{ij} \dot\za_i \dot\zb_j ~,
\quad \Psi\left( X , Y \right) = \frac 1 2 \sum_{i,j=1}^n L_{ij} X_i X_j
\ee
Which characterize the real evolution only when $\dot\za=\dot\zb$ are the real evolving
fluxes and when $X=Y$ are the real thermodynamic forces, in which cases we have:
\be
\dot{\cal S} = 2 \Phi\left( \dot \za , \dot \za \right) = 2 \Psi\left( X , X \right)
\ee
The molecular chaos may be accounted for by a random perturbation, which turns Eq.(\ref{hydrod}) into
\be
\sum_{j=1}^n \left[ R_{ij}\dot{\za}_j + s_{ij} \za_j \right] = \ze_i ~, \quad \langle \ze_i \rangle=0~,
\qquad i=1,...,n
\label{randhyd}
\ee
where $\ze_i$ is a random force which allows different
paths with different probabilities and which does no net work.

Let $f_{i,1}\left(\za_i^{(1)} , t_1 \right)$ be the probability density for the $i$-th observable to
take values close to $\za_i^{(1)}$ at time $t_1$. By assumption {\bf A3}, $f_{i,1}$ is independent of $t_1$. Let
$f_{i,1} \left( \za_i^{(k)} , t_k | \za_{i-1}^{(k-1)} , t_{k-1} \right)$ be the conditional probability
density for the $i$-th observable to take values close to $\za_i^{(k)}$ at time $t_k$, given that
it was $\za_{i-1}^{(k-1)}$ at time $t_{k-1}$. Because of the Markov property and of {\bf A3}, one has:
\bea
\hskip -5cm
&& \hskip -1cm f_{i,p}\left(\za_i = \za_i^{(k)} , t_k,  k=1,...,p\right) \\
&& \hskip -1cm =
f_{i,1} \left( \za_i^{(p)} , t_p | \za_{i-1}^{(p-1)} , t_{p-1} \right)
\cdots f_{i,1} \left( \za_i^{(2)} , t_2 | \za_{i}^{(1)} , t_{1} \right)
f_{i,1}\left(\za_i^{(1)} , t_1 \right) \\
&& \hskip -1cm =
f_{i,1} \left( \za_i^{(p)} , t_p | \za_{i-1}^{(p-1)} , t_{p-1} \right)
\cdots f_{i,1} \left( \za_i^{(2)} , t_2 | \za_{i}^{(1)} , t_{1} \right)
e^{{\cal S}(\za^{(1)})/k_{_B}}
\label{onsg}
\eea
with two constraints
\be
\mbox{ a) } \lim_{\zt \to 0} f_{i,1} \left( \za_i , t_1+\zt | \za_{i}^{(1)} , t_{1} \right) =
K \zd \left(\za -\za^{(1)} \right)
\ee
due to the fact that $\zt \to 0$ is the limit in which $\za$ deterministically approaches $\za^{(1)}$, and
\be
\mbox{b) } \lim_{\zt \to \infty} f_{i,1} \left( \za_i , t_1+\zt | \za_{i}^{(1)} , t_{1} \right) =
e^{{\cal S}(\za^{(1)})/k_{_B}}
\ee
representing the loss of correlations between the time $t_1$ and the time $t_1+\zt$. Solving
the Langevin equation (\ref{randhyd}), $f_{i,1} \left( \za_i , t_1+\zt | \za_{1}^{(1)} , t_{1} \right)$
can be explicitly given.
Let us now turn to the case with $n=1$:
\be
R \dot\za + s \za = \ze
\ee
this process is described by:
\be
f_1 \left( \za , t+u | \za^{(0)} , t \right) =
\frac{s \exp\left\{ -\frac{s\left(\za - \za^{(0)} e^{-su/R} \right)^2}{2k_{_B}
\left(1-e^{-2su/R}\right) }\right\}}
{\sqrt{2\pi} k_{_B} \sqrt{1-e^{-2su/R}}}
\ee
With this information and with Ito's discretization convention \cite{OM53a},
one eventually obtains:
\bea
&&f_1 \left( \za , t+\zt | \za^{(0)} , t \right) =
\left(\frac 1{2k_{_B}}\right)^p \left(\frac{sR}{\pi \zd\zt}\right)^{p/2} \times \\
&&\hskip 25pt \int {\rm d} \za^{(1)} \cdots \int {\rm d} \za^{(p)}
\exp \left\{ -\frac{R}{4k_{_B}} \sum_{k=1}^p
\left[ \dot\za^{(k)}+\frac s R \za^{(k+1)} \right]^2 \zd \zt \right\} \\
\eea
Under the $p \to \infty$, $\zd\zt \to 0$  limits, with $\tau = p \zd\zt$, the sum in the
exponential tends to the integral along the path:
\be
\int_t^{t+\zt} \left[ \dot\za(t')+\frac s R \za(t') \right]^2 {\rm d}t'
\ee
which must be minimized to maximize the probability. Analogously, the $n$-dimensional
case requires the minimization of:
\be
\int_t^{t+\zt} \sum_{i=1}^n \left[ \dot\za_i(t')+\frac{s_i} R_i \za_i(t') \right]^2 {\rm d}t' .
\ee
Here, the integrand can be expressed as
\be
{\cal L}\left( \za , \dot\za \right) = 2 \Phi\left( \dot\za , \dot\za \right)
-2 \dot{\cal S}(\za) + 2 \Psi\left( X(\za) , X(\za) \right)
\ee
and the path of minimum integral follows from the Lagrange equation:
\be
\frac{\rm d}{{\rm d} t} \frac{\partial {\cal L}}{\partial \dot\za} -
\frac{\partial {\cal L}}{\partial \za} = 0 ~, \quad \mbox{which yields} ~~~
R_j \ddot{\za}_j - \frac{s_j^2}{R_j} \za_j = 0 ~, \quad j=1,...,n
\ee
These second order differential equations are equivalent to
pairs of first order equations. Indeed, their general solution
\be
\za_j(t) = C_{j1}~e^{-s_j t/R_j} + C_{j2}~e^{s_j t/R_j}
\ee
requires $C_{j2}=0$ when the $t \to \infty$ limit is considered --in which case we have
 relaxation to equilibrium from a nonequilibrium initial condition-- while it requires
$C_{j1}=0$ when the previous history, beginning with an
equilibrium state at $t= -\infty$, is
considered. The first case is solution of the
differential equation
\be
\dot\za_j + \frac{s_j}{R_j} \za_j = 0
\ee
and the second case corresponds to
\be
\dot\za_j - \frac{s_j}{R_j} \za_j = 0 ~.
\ee
We thus have two evolutions, which are symmetric under time reversal:
one describes the relaxation to equilibrium, in accord with hydrodynamics;
the other treats fluctuations away from equilibrium, and is the first
example of the so-called {\em adjoint hydrodynamics} \cite{Jona}. In the large $n$
limit, the most probable path
becomes the only path of positive probability and
a justification of hydrodynamics is obtained, starting from a mesoscopic
description.

\vskip 5pt
\noindent
{\em These results are crucially based on the
Gaussian distributions, hence they are restricted to small deviations, from which
the linear response about equilibrium states is derived.}

\vskip 5pt
Considering large deviations, this theory has been generalized
to fluctuations about nonequilibrium steady states, which are not
symmetric under time reversal \cite{Jona}.
For dissipative deterministic particle systems, that are time reversal invariant, it has
been shown that similar asymmetries may arise, when particles interact \cite{GRV}.

\section{Fluctuation Relations: response from large deviations}
\label{sec4}
In 1993, the paper \cite{ECM} addressed the question of the fluctuations of the entropy
production rate in a pioneering attempt towards a unified theory of a wide range of
nonequilibrium phenomena. In particular, a {\em Fluctuation Relation} (FR) was there
derived and tested. Obtained on purely dynamical grounds, it constitutes one of the few
general exact results for systems almost arbitrarily
far from equilibrium, while close to equilibrium it is consistent with the Green-Kubo
and Onsager relations.
This FR reads:
\begin{equation}
\frac{{\rm Prob}_\zt(\zs \approx A)}{{\rm Prob}_\zt(\zs \approx -A)} = e^{\zt A}
\label{firstFR}
\end{equation}
where $A$ and $-A$ are average values of the normalized power dissipated in a long time $\zt$ in a driven system,
denoted by $\zs$ and ${\rm Prob}_\zt(\zs \approx \pm A)$ is the steady state
probability of observing values close to $\pm A$.

\vskip 5pt
\noindent
{\em This relation constitutes a {\em large deviation} result: for large $\zt$, any $A \ne \langle \zs \rangle$
lies many standard deviations away from the mean. In other words, A
corresponds to a large (macroscopic) deviation from the macroscopically observable
value $\langle \zs \rangle$.}

\vskip 5pt
The FR (\ref{firstFR}) was derived for the following {\em isoenergetic} model of a 2-dimensional
shearing fluid:
\begin{equation}
\left\{
\begin{array}{l}
 \dfrac{d}{d t}{\bf q}_i = \dfrac{{\bf p}_i}{m} + \gamma \, y_i {\bf \hat{x}}  \\
 \\
 \dfrac{d}{d t}{\bf p}_i =  {\bf F}_i ({\bf q}) + \gamma  p_i^{(y)} {\bf \hat{x}} -\alpha_{th} {\bf p}_i
\end{array}
\right.
\label{SLLODeqs}
\end{equation}
where $\gamma$ is the shear rate in the $y$ direction, ${\bf \hat{x}}$ is the unit vector in
the $x$-direction, and the friction term $\za_{th}$, called ``thermostat'', takes the form
\begin{equation}
\alpha_{th} (\zG) = - \frac{\zg}{\sum_{i=1}^N {\bf p}_i^2} ~
\sum_{i=1}^N p^{(x)}_i p^{(y)}_i \label{IKsllalpha}
\end{equation}
\noindent
as prescribed by Gauss' principle of least constraint, in order to keep the internal
energy fixed.

This molecular dynamics model was chosen by the authors of \cite{ECM} because its phase
space expansion rate $\zL$ is proportional to $\za_{th}$. Hence a dynamical quantity,
could be related to the energy dissipation rate divided
by $\sum {\bf p}_i^2$. The FR is parameter-free and, being dynamical in nature, it applies
almost arbitrarily far from equilibrium as well as to small systems.

Gallavotti and Cohen clearly identified the mathematical framework within which Ref. \cite{ECM} had been developed,
 introducing the following
\cite{GCa,GCb,GG-MPEJ,GGrevisited}:

\vskip 5pt \noi {\bf Chaotic Hypothesis:} {\it A reversible many-particle
system in a stationary state can be regarded as a transitive Anosov system for
the purpose of computing its macroscopic properties.}

\vskip 5pt \noi
Anosov systems can indeed be proven to have probability distributions
of the kind assumed in \cite{ECM}. The result is a steady state FR for the fluctuations
of $\zL$, which we call $\zL$-FR and which will be described below.
As the Anosov property practically means a high degree of randomness, analogous results
have been obtained first for finite state space Markov chains
and later for many other stochastic
processes \cite{Kurchan,Lebowitz-Spohn,Maes}. Stochastic processes are easier to
handle than deterministic dynamics, but ambiguities affect their observables, except
for special cases. The reader is addressed to the
numerous existing review papers, such as Refs.\cite{MeRo07,BPRV,Gaw}. We focus
now on some specific results for deterministic dynamics.

\subsection{The Gallavotti-Cohen approach}
\label{GCapproach}
The idea proposed by Gallavotti and Cohen is that dissipative,
reversible, transitive Anosov maps, $S : \mathcal{M} \to \mathcal{M}$, are idealizations of nonequilibrium
particle systems \cite{GCb}. That the system evolves with discrete or continuous time was
thought to be a side issue \cite{GCb}. The $\zL$-FR for
Anosov maps relies on time reversibility and on the fact that these dynamical systems
admit arbitrarily fine \emph{Markov} partitions \cite{sinaibook}. These are subdivisions
of $\mathcal{M}$ in cells with disjoint interiors and with boundaries forming
invariant sets, which in two dimensions consist of pieces of stable
and unstable manifolds.
Gallavotti and Cohen further assumed that the dynamics is transitive, \ie that
a typical trajectory explores all regions of $\mathcal{M}$, as finely as one wishes.
This structure justifies the probability (Lyapunov) weights of
Eq.(1) in Ref.~\cite{ECM}, from which the \LFR emerges.

Let the dynamics be given by $X_{k+1}=SX_{k}$ and introduce the phase
space expansion rate $\zL(X) = \log J(X)$, where $J$ is the Jacobian determinant
of $S$. The dynamics is called {\em dissipative} if $\langle \zL \rangle < 0$, where
$\langle . \rangle$ is the steady state phase space average.
Then, consider the dimensionless phase space
contraction rate $e_\zt$, obtained
along a trajectory segment $w_{X,\zt}$ with origin at $X \in \mathcal{M}$ and
duration $\zt$, defined by:
\begin{equation}
e_\zt(X) = \frac{1}{\zt \langle \zL \rangle} \sum_{k=-\zt/2}^{\zt/2-1} \zL(S^k X)
\label{p}
\end{equation}
Let $J^u$ be the Jacobian determinant of $S$ restricted to the unstable manifold $V^+$,
\ie the product of the asymptotic separation factors of nearby points, along the
directions in which distances asymptotically grow at an exponential rate.
{\em If the system is Anosov}, the probability that
$e_\zt(X) \in B_{p,\ze}\equiv(p-\ze,p+\ze)$ equals, in the fine Markov partitions
and long $\zt$ limits, with the sum of weights of form
\be
w_{X,\zt}=\prod_{k=-\zt/2}^{\zt/2-1} \frac{1}{J^u(S^k X)}
\ee
of the cells containing the points $X$ such that $e_\zt(X) \mbox{ lies in } B_{p,\ze}$.
Then, denoting by $\pi_{\tau}(B_{p,\ze})$ the corresponding probability, one can write
\begin{equation}
\pi_\zt(e_\zt(X) \in B_{p,\ze}) \approx \frac{1}{M_\zt} \sum_{X: e_\zt(X)\in B_{p,\ze}} w_{X,\zt}
\label{Piofp}
\end{equation}
where $M_\zt$ is a normalization constant.
{\em If the support of the physical measure is} $\mathcal{M}$, as in the case
of moderate dissipation  \cite{ECSB}, time-reversibility and dissipation
guarantee that the range of possible fluctuations includes a symmetric interval
$[-p^*,p^*]$, with $p^*>0$, and one can consider the ratio
\begin{equation}
\frac{\pi_\zt(B_{p,\ze})}{\pi_\zt(B_{-p,\ze})} \approx
\frac{\sum_{X,e_\zt(X) \in B_{p,\ze}} w_{X,\zt}}
{\sum_{X,e_\zt(X)\in B_{-p,\ze}} w_{X,\zt}} ~,
\label{pminusp}
\end{equation}
where each $X$ in the numerator has a counterpart in the denominator.
Denoting by $I$ the involution which replaces the initial condition of a
given trajectory with the initial condition of the reversed
trajectory, time-reversibility yields:
\begin{equation}
\zL(X)=-\zL(IX) ~, \quad w_{IX,\zt} = w_{X,\zt}^{-1} \quad \mbox{and~~~ }
\frac{w_{X,\zt}}{w_{IX,\zt}} = e^{-\zt \langle \zL \rangle p}
\end{equation}
\\
if $e_\zt(X)=p$.
Taking small $\ze$ in $B_{p,\ze}$, the division of each term in the numerator
of (\ref{pminusp}) by its counterpart in the denominator approximately equals
$e^{-\zt \langle \zL \rangle p}$, which then equals the ratio in
(\ref{pminusp}). Therefore, in the limit of small $\ze$, infinitely fine Markov partitions
and large $\zt$, one obtains the following:

\vskip 5pt\noindent
{\bf Gallavotti-Cohen Theorem.} {\em Let $(\mathcal{M},S)$ be dissipative and reversible
and assume that the chaotic hypothesis holds. Then, in the $\zt \to \infty$ limit, one
has
\begin{equation}
\frac{\pi_{\tau}(B_{p,\ze})}{\pi_{\tau}(B_{-p,\ze})} = e^{- \zt \langle \zL \rangle p} ~.
\label{largedev}
\end{equation}
with an error in the argument of the exponential which can be
estimated to be $p$- and $\zt$-independent.}
\vskip 5pt

\noindent
If $\zL$ can be identified with a physical observable, the \LFR
is a parameter-free statement about the physics of nonequilibrium systems.
Unfortunately, $\zL$ differs from the dissipated power in general, \cite{ESR},
hence alternative approaches have been developed.

\subsection{Fluctuation relations for the dissipation function}
\label{sec:ESR}
{One different approach from above consists in posing a different question in order to remain closer to the
interest of physics:
if the FR has been observed to hold for the energy dissipation
of a given system, which mechanisms are responsible for that?}
To answer this question, various results have been achieved and others clarified.
In particular:
\begin{itemize}
\item[\bf 1.] transient, or ensemble, FRs have been derived;
\item[\bf 2.] classes of infinitely many identities have been obtained to characterize
equilibrium and nonequilibrum states;
\item[\bf 3.] a novel ergodic notion, known as {\em t-mixing}, has been introduced;
\item[\bf 4.] a quite general response formula has been derived.
\end{itemize}

These develoments began with a paper by Evans and Searles \cite{earlierpapersA},
who proposed the first transient fluctuation relation for the {\em Dissipation  Function}
$\zW$, which is formally similar to Eq.(\ref{firstFR}). In
states close to  equilibrium, $\zW$ can be  identified with  the {\em
entropy production rate}, $\zs = J V  F^{ext} / k_{_B} T$, where, $J$ is the
(intensive) flux due  to the thermodynamic force $F^{ext}$,  $V$ and $T$ are the volume
and the  kinetic temperature, respectively  \cite{earlierpapersA,earlierpapersB}. This
relation,  called   transient  $\zW$-FR,  is  obtained under  virtually  no
hypothesis, except  for {\em time  reversibility}; it is transient because it
concerns non-invariant ensembles of systems, instead of the steady state.
The approach stems from the belief  that the complete knowledge of the
invariant  measure  implied  by  the  Chaotic  Hypothesis  is  not  required  to
understand the few properties of physical interest, like thermodynamic relations
do not  depend on the details of the microscopic dynamics \cite{ESR2}.

Let ${\cal M}$ be the phase space of the system at
hand, and $S^\tau: {\cal M} \rightarrow {\cal M}$ be a reversible evolution corresponding to
$\dot \zG = F (\zG)$. Take a  probability measure $d \mu_0(\zG) = f_0(\zG) d \zG$
on ${\cal M}$, and let the observable $\mathcal{O} : {\cal M} \rightarrow \zR$
be  odd with respect to the time reversal,  {\it i.e.}  \ $\mathcal{O}(I  \zG) =
-\mathcal{O}(\zG)$. Denote its time averages by
\begin{equation}
\Ft(\zG) \equiv \frac{1}{\tau} \mathcal{O}_{t_0,t_0+\zt}(\zG) \equiv
 \frac{1}{\tau}
\int_{t_0}^{t_0+\tau} \mathcal{O}(S^{s} \zG) d s ~.
\label{phitau}
\end{equation}
For a  density $f_0$ that is even under time reversal [$f_0(I \zG)=f_0(\zG)$],
define the \vskip 5pt

\noi
{\bf  Dissipation function:}
\bea \label{omegat}
&&\zW(\zG)  = - \left. \frac{d}{d \zG} \ln f_0
\right|_\zG   \cdot  \dot{\zG} - \zL(\zG)  ~, \quad \mbox{so that} \\
&&\Wt(\zG) = \frac{1}{\zt}
\left[ \ln \frac{f_0(S^{t}\zG)}{f_0(S^{t+\zt} \zG)} -
\zL_{t,t+\tau} \right] \label{integrW}
\eea
For  a   compact   phase   space,  the  uniform   density
$f_0(\zG)=1/|{\cal M}|$ implies $\zW=\zL$, which was the case of the original FR.
The existence of the logarithmic term in
(\ref{omegat})  is  called  {\em ergodic  consistency},  a
condition met if $f_0>0$ in all regions visited by all trajectories $S^t \zG$.

For $\zd > 0$,  let $A^\pm_\zd=(\pm A-\zd,\pm A+\zd)$, and
let  $E(\mathcal{O}  \in  (a,b))$  be  the  set  of  points  $\zG$ such  that
$\mathcal{O}(\zG) \in (a,b)$.   Then, we have $E(\Wz \in A^-_\zd) =  I S^\zt E(\Wz \in
A^+_\zd)$ and:
\bea
\hskip -13pt\frac{\mu_0(E(\Wz \in A^+_\zd))}{\mu_0(E(\Wz \in A^-_\zd))}
&=&\frac{ \int_{E(\Wz \in A^+_\zd)} f_0 (\zG) d \zG }{
\int_{E(\Wz \in A^+_\zd)}
f_0(S^\zt X) e^{-\zL_{0,\zt}(X)} d X } \nonumber \\
&=&\frac{ \int_{E(\Wz \in A^+_\zd)} f_0(\zG) d \zG }{
\int_{E(\Wz \in A^+_\zd)} e^{-\zW_{0,\zt}(X)} f_0(X) d X }
=\left\langle e^{-\zW_{0,\zt}} \right\rangle_{\Wz \in
  A^+_\zd}^{-1} \nonumber
\label{ESFRp}
\eea
where by $\left\langle \cdot \right\rangle_{\Wz  \in A^+_\zd}$
we mean the average computed with respect to $\mu_0$ under the condition that $\Wz  \in A^+_\zd$. This implies
the
\vskip 5pt
\noindent
{\bf Transient $\zW$-FR:}
\begin{equation} \label{ESFR}
\frac{\mu_0(E(\Wz \in A^+_\zd))}{\mu_0(E(\Wz \in A^-_\zd))} =
e^{[A+\ze(\zd,A,\zt)]\zt} \ ,
\end{equation}
with  $|\ze(\zd,A,\zt)|  \le  \zd$, an  error due  to the  finiteness
of  $\zd$.

\vskip 5pt
\noindent
{\bf Remarks:} {\em
\begin{itemize}
\item[{\bf i.}] The transient $\zW$-FR refers to the non-invariant probability
distribution $\mu_0$. Time reversibility is basically the only ingredient of  its derivation.
\item[{\bf ii.}] Its similarity with the steady state FR is
misleading: rather than expressing a statistical property of fluctuations of
a given system, it expresses a property of the initial ensemble of macroscopically identical systems.
\item[{\bf iii.}] In order for $\zW$ to be the energy dissipation, $f_0$ has to be properly chosen.
For instance, in simple molecular dynamics models, $\Omega$ is the energy dissipation if $f_0$ is the equilibrium
ensemble dynamics, which is obtained when the
external driving is switched off, while the thermostats keep acting.
\item[{\bf iv.}] Consequently, {\em the transient $\zW$-FR yields a property of the equilibrium state
by means of nonequilibrium experiments}, thus complementing the FDR, which yields non equilibrium properties
from equilibrium experiments.
\end{itemize}} \noi
The  steady state $\zW$-FR requires further hypotheses.
In the first place let averaging begin at time $t$, i.e.\ consider
\begin{equation}
\frac{\mu_0(E(\Wt \in A^+_\zd)) }{\mu_0(E(\Wt \in A^-_\zd)) } \ .
\label{PpoverPp1}
\end{equation}
Taking $\hat t=t +\zt+t$, the transformation $\zG=IS^{\hat t} W$ in $\cal{M}$ and some algebra yield:
\bea
\frac{\mu_0(E(\Wt \in A^+_\zd)) }{\mu_0(E(\Wt \in A^-_\zd)) } &=&
\left\langle \exp \left( - \zW_{0,\hat t} \right)
\right\rangle_{\overline{\zW}_{t,t+\zt} \in A^+_\zd}^{-1} \\
&=& e^{\left[ A + \ze(\zd,t,A,\zt)\right] \zt}
\left\langle e^{- \zW_{0,t} -  \zW_{t+\zt,2t+\zt}}
\right\rangle_{\Wt \in A^+_\zd}^{-1}
\label{FtWt}
\eea
where $|\ze(\zd,t,A,\zt)| \le \zd$. Here, the second line follows from the first because
$\zW_{0,\hat t}=\zW_{0,t}+\zW_{t,t+\zt}+\zW_{t+\zt,\hat t}$, with the central contribution
made approximately equal to $A$ by the condition $\Wt \in A^+_\zd$.
Recall that $\mu_0(E) = \mu_{t}(S^{t} E)$, where $\mu_{t}$
is the evolved probability distribution, with density
$f_{t}$. Then, taking the logarithm and dividing by
$\zt$ Eq.(\ref{FtWt}) produces:
\bea
&&\frac{1}{\zt} \ln
\frac{\mu_{t}(E(\overline{\zW}_{0,\zt} \in A^+_\zd))}
{\mu_{t}(E(\overline{\zW}_{0,\zt} \in A^-_\zd))} = \nonumber \\
&&\hskip 45pt =A + \ze(\zd,t,A,\zt)
- \frac{1}{\zt} \ln
\left\langle e^{-\zW_{0,t} - \zW_{t+\zt,2t+\zt}} \right\rangle_{\Wt \in A^+_\zd}
\label{SSESFT} \\
&&\hskip 45pt \equiv A + \ze(\zd,t,A,\zt) + M(A,\zd,t,\zt) \nonumber
\eea
because {$E(\overline{\zW}_{0,\zt})=S^tE(\Wt)$}.

If $\mu_{t}$  tends to  a steady state  $\mu_\infty$ when $t  \to \infty$, the
exact relation (\ref{SSESFT}) changes from a statement on the ensemble $f_{t}$, to
a statement on the statistics generated  by a single typical trajectory.
In particular one could have the analogous of the $\Lambda$-FR:

\vskip 5pt  \noi {\bf Steady State  $\zW$-FR.} {\it For  any tolerance $\ze>0$, there
is a sufficiently small $\zd > 0$
such that
\begin{equation}
\lim_{\zt \to \infty}
\frac{1}{\zt} \ln
\frac{\mu_{\infty}(E(\overline{\zW}_{0,\zt} \in A^+_\zd))}
{\mu_{\infty}(E(\overline{\zW}_{0,\zt} \in A^-_\zd))} = A + \eta ~, \quad \mbox{with }~
\eta \in (-\ze,\ze)
\label{SSFTestim}
\end{equation}
}
\vskip 5pt \noi For this to be the case, one needs some assumption.
Indeed, $M(A,\zd,t,\zt)$ could diverge with $t$ at fixed $\tau$, making Eq. (\ref{SSESFT}) useless.
If on the other hand $M(A,\zd,t,\zt)$ remains bounded
by a finite  $M(A,\zd,\zt)$, $\lim_{\zt \to  \infty} M(A,\zd,\zt)$
could still exceed $\ze$.

The first difficulty is simply solved
by  the  observation  that  the  divergence of  $M(A,\zd,t,\zt)$  implies
that one of the probabilities on the left hand side of Eq.(\ref{SSESFT}) vanishes,
i.e./ that $A$ or $-A$ are not observable in the steady state.
If no value $A$ is observable, there are no fluctuations in the steady state and
there is no need for a steady state FR. Therefore,  let us  assume
that $A$ and $-A$ are observable.
To proceed, observe that Eqs.(\ref{omegat},\ref{integrW}) lead to
\be
f_s (\zG) = f_0\left(S^{-s} \zG \right) e^{-\zL_{-s,0}(\zG)} = f_0(\zG) e^{\zW_{-s,0}(\zG)}
\label{fsevol}
\ee
which implies the following relation:
\begin{equation}
\left\langle e^{-\zW_{0,s}} \right\rangle_0 = 1 ~, \quad \mbox{for every } s \in \zR ~.
\label{normalizat}
\end{equation}
Suppose now that the $\zW$-autocorrelation with respect to $f_0$ decays instantaneously in time, so
that one can write:
\begin{equation}
1 = \left\langle e^{-\zW_{0,s} -\zW_{s,t} } \right\rangle_0 =
\left\langle e^{-\zW_{0,s}} \right\rangle_0 \left\langle e^{-\zW_{s,t}} \right\rangle_0 ~,
\end{equation}
hence
\be
\left\langle e^{-\zW_{s,t}} \right\rangle_0 =1 ~, \quad \mbox{for all $s$ and $t$}
\ee
under the same condition, the conditional average of eq.(\ref{SSESFT})
does not depend on the condition ${\Wt\in A^+_\zd}$, so that:
\begin{equation}
\left\langle e^{-\zW_{0,t}} \cdot e^{-\zW_{t+\zt,2t+\zt}}
\right\rangle_{\Wt\in A^+_\zd} =
\left\langle e^{-\zW_{0,t}} \cdot e^{-\zW_{t+\zt,2t+\zt}} \right\rangle_0
= 1 ~.
\label{condave-full}
\end{equation}
Then, the logarithmic correction  in Eq. (\ref{SSESFT}) identically vanishes
for all $t,\zt$, and the steady state \WFR is  verified at all $\zt > 0$. This
idealized  situation does  not need to be realized, but
molecular dynamics indicates that the typical situation is
similar to this \cite{ESR3}; for $\tau$ much larger than a characteristic time $\tau_M$,
one may write:
\begin{eqnarray}
&&\left\langle
e^{-\zW_{0,t}} \cdot e^{- \zW_{t+\zt,2t+\zt}}
\right\rangle_{\overline{\zW}_{t,t+\zt} \in A_\zd^+} \approx \label{prim}\\
&&\hskip 55pt\approx \left\langle e^{-\zW_{0,t-t_M}} \cdot e^{- \zW_{t+\zt+t_M,2t+\zt}}
\right\rangle_{\overline{\zW}_{t,t+\zt} \in A_\zd^+} \ \\
&&\hskip 55pt\approx \left\langle e^{-\zW_{0,t-t_M}} \cdot e^{- \zW_{t+\zt+t_M,2t+\zt}}
\right\rangle_0 \\
&&\hskip 55pt\approx \left\langle e^{-\zW_{0,t+t_M}} \right\rangle_0
\left\langle e^{- \zW_{t+\zt+t_M,2t+\zt}} \right\rangle_0 = O(1) \ ,
\label{newargdeco}
\end{eqnarray}
with improving accuracy for growing $t$ and $\zt$.
If  these  scenarios  are  realized,
$M(A,\zd,\zt)$ vanishes  as $1/\zt$ for growing $\tau$.
\vskip 5pt
\noindent
{
The assumption that Eqs.(\ref{prim})-(\ref{newargdeco}) hold is a kind of mixing property which, however, refers to non-invariant probability
distributions, differently from the standard notion of mixing.}
\vskip 5pt
\noindent
Various other relations can be obtained following the same procedure. For instance,
for each odd $\mathcal{O}$, any $\zd>0$, any $t$ and any $\zt$ the following transient
FR holds:
\begin{equation}
\frac{\mu_{0}(\overline{\mathcal{O}}_{0,\zt} \in A_\zd^+)}
{\mu_{0}(\overline{\mathcal{O}}_{0,\zt} \in A_\zd^-)} =
\left\langle \exp \left( -\Omega_{0,\zt} \right) \right\rangle^{-1}_{\Fz \in A_\zd^+}\ ,
\label{inter}
\end{equation}
expressed a property of the initial
state by means of nonequilibrium dynamics.

\section{t-mixing and general response theory}
\label{sec5}

\vskip 5pt
\noindent
Observing that Eq.(\ref{useful}), implies:
\be
\left\langle e^{- \zW_{s,t}} \right\rangle_0 = \left\langle e^{- \zW_{0,t-s}} \right\rangle_s
\ee
Eqs.(\ref{prim}-\ref{newargdeco}) appear to be one
special case of the following property:
\be
\lim_{t \to \infty} \left[
\left\langle \psi \left( \phi \circ S^t \right) \right\rangle_0
- \left\langle \psi \right\rangle_0 \left\langle \phi \right\rangle_t \right] = 0
\label{1tmi}
\ee
\\
In the case that $\psi = \zW$, Eq.(\ref{1tmi}) becomes
\be
\lim_{t \to \infty}
\left\langle \zW \left( \phi \circ S^t \right) \right\rangle_0 = 0
\ee
because $\zW$ is odd and $f_0$ is even under time reversal, hence $<\Omega>_0=0$.\\
If the convergence of this limit is faster than $O(1/t)$, one further has:
\be
\int_0^\infty \left\langle \zW \left( \phi \circ S^t \right) \right\rangle_0 ~
{\rm d} t \in \mathbb{R}
\label{tmixin}
\ee
a condition which has been called {\em t-mixing}.

To obtain the response of observables, starting from an equilibrium
state, we have:
\be
\langle \phi \rangle_t - \langle \phi \rangle_0 = \int_0^t
\frac{\rm d}{{\rm d} s} \langle \phi \rangle_s~ {\rm d} s =
\int_0^t  {\rm d} s \frac{\rm d}{{\rm d} s} \int {\rm d} \zG f_s(\zG) \phi(\zG)
\ee
Where Eq.(\ref{fsevol}) yields:
\be
\frac{\rm d}{{\rm d} s} \int {\rm d} \zG f_s(\zG) \phi(\zG) =
\int  {\rm d} \zG f_0(\zG) e^{\zW_{-s,0}(\zG)} \zW\left(S^{-s}\zG\right) \phi(\zG)
\ee
Introducing the coordinate change $X=S^{-s}\zG$, $\zG = S^s X$, with Jacobian determinant
$|\partial \zG/\partial X|=\exp(\zL_{0,s}(X)$ and observing that:
\be
\zW_{-s,0}(S^sX) = \int_{-s}^0 {\rm d} u~\zW\left(S^u S^s X \right)
= \int_0^s {\rm d} z~\zW(S^zX) = \zW_{0,s}(X)
\ee
so we finally obtain:
\bea
\frac{\rm d}{{\rm d} s} \left\langle \phi(\zG) \right\rangle_s &=&
\int {\rm d} X~ \phi\left(S^sX\right) \zW(X) e^{\zW_{0,s}(X)} e^{\zL_{0,s}(X)}f_0(S^sX) \\
&=& \int {\rm d} X~\zW(X) \phi\left(S^sX\right) f_0(X) =
\left\langle \zW \left( \phi \circ S^s \right) \right\rangle_0
\eea
which is the integrand of Eq.(\ref{tmixin}). Therefore, we have the following
Response Formula:
\be
\langle \phi \rangle_t = \langle \phi \rangle_0 +
\int_0^t {\rm d} s~ \left\langle \zW \left( \phi \circ S^s \right) \right\rangle_0
\label{genresp}
\ee
Moreover, if the t-mixing condition holds for $\phi$, we get
\be
\langle \phi \rangle_t \stackrel{t \to \infty}{\longrightarrow}
\langle \phi \rangle_0 +
\int_0^\infty {\rm d} s~ \left\langle \zW \left( \phi \circ S^s \right) \right\rangle_0 \in \mathbb{R}
\ee
and the ensemble under investigation converges to what appears to be a steady state.

One interesting aspect of the relation between standard mixing and t-mixing is the following.
Standard mixing concerns the decay of correlations among the
evolving microscopic phases within a given steady state, t-mixing concerns the decay of
correlations among evolving macrostates. For this reason, the t-mixing property implies
the convergence to a steady state, whereas the mixing property in general does not.

Mixing assumes the state to be stationary, making
irrelevant the issue of relaxation. The derivation of
convergence to a microcanonical state, illustrated in Section \ref{ErgMix}, is thus
just a trick.
That derivation is possible because one may formally interpret the evolving transient probability
densities as evolving {\em observables} as well. This way one combines in one mathematical
object two physically very different entities: the ensemble of microscopic phases and
a macroscopic measurable observable.\footnote{Something similar happens
when the equilibrium
thermodynamic entropy of a physical object is expressed by the equilibrium average of
the logarithm of the equilibrium density, which is the Gibbs entropy.}
This will not be legitimate under most circumstances. However, even in the case of $t$-mixing,the convergence of the
steady state has not been proved in the sense of thermodynamics. Indeed, different initial conditions $\Gamma \in
\mathcal{M}$ are allowed by $t$-mixing to produce different time averages. The uniqueness of the time average is
currently under investigation.

\section{Stochastic diffusions and large deviations}
\label{sec6}

Let us now turn our attention to stochastic dynamics. In general, the presence of noise allows
one to characterize the steady state dynamics, even in presence of dissipation, by
regular probability densities, thus overcoming the problem posed e.g.\ by fractal structures.
Hence, one may safely rely, in this case, on perturbative approaches in the description of perturbations of a given
(possibly dissipative) reference state. In particular, a detailed analysis of the response formulae valid for Markovian
Langevin-type stochastic differential equations is presented in Ref. \cite{R4}, where Ruelle clarifies
the conditions under which the zero noise limit leads the various terms of the perturbation theory to reproduce
theie counterparts in the deterministic dynamics, cf. Refs. \cite{R1,mattval}.
In Ruelle's case, this is made possible
by the stability of the SRB states under small random perturbations \cite{Young, Kifer}. \\
A different approach based on the large deviations method is presented in Refs. \cite{CMW, Baies}.
Ler us focus, for simplicity, on stochastic diffusion processes described by
overdamped Langevin equations, in which one disregards inertial effects, letting
forces to be proportional to velocities rather than to accelerations
\cite{Maes1,MNW}. These processes correspond to the high damping limits of the underdamped (or \textit{inertial}) stochastic dynamics.
Let us start considering overdamped diffusion processes for $x\in {\mathbb R}^n$, in the It\^o sense expressed by:
\be
\dot{x}_t = \chi \cdot [F(x_t)+ F^p_t(x_t)] + \nabla \cdot D(x_t)+\sqrt{2
D(x_t)}\,\xi_t \quad , \label{diff}
\ee
where $\xi_t$ denotes standard white noise and $F^p_t$ denotes the perturbation to the reference dynamics.
The mobility $\chi$ and the diffusion constant $D$ are strictly positive (symmetric)
$n\times n$-matrices, which, provided the system is in contact with a
thermostat at inverse temperature $\beta>0$, are connected by the
Einstein relation $\chi=\beta D$.
The force $F$ denotes the drift of the reference unperturbed dynamics, and can be expressed as:
\be
F = F_{nc}- \nabla U  \label{force} \quad ,
\ee
where $F_{nc}$ denotes a nonconservative force pulling the reference dynamics out of equilibrium, while $U$ is the energy of the system.
The Fokker-Planck equation for the time dependent density $f_t$, related to the diffusion process described by (\ref{diff}), reads
\be
 \frac{\partial f_t}{\partial t}(x_t) = -\nabla \cdot j_{f} \hskip 2pt, \hskip 8pt
\mbox{with} \hskip 8pt j_{f}=[\chi (F+ F^p_t)f_t(x_t)-\frac{\chi}{\beta}\nabla f_t(x_t)] \label{fp}   \quad ,
\ee
where $j_{f}$ denotes the probability current \cite{risken}.
Rather than attempting a direct solution of Eq. (\ref{fp}), one may tackle Eq. (\ref{diff}) from
the point of view of large deviations theory \cite{CMW,dembo}.
The key idea, cf. Refs. \cite{CMW,MN}, is to determine the perturbed probability density through its embedding
in the path-space distribution. That is, given the (random) paths $\omega = (x(s),
s\in [0,t])$, one may connect the distribution $P$ on paths starting from $f_0$
and subjected to the perturbation $F^p_t$, with the reference distribution $P^o$
pertaining to paths starting from $f_0$ and undergoing the reference dynamics, via
the formula:
\begin{equation}\label{act}
P(\omega) =e^{-\mathcal{A}(\omega)}\,P^o(\omega) \quad .
\end{equation}
The relation (\ref{act}) defines the action $\mathcal{A}(\omega)$, which is typically local in
space-time and is, thus, similar to the Hamiltonians or Lagrangians of
equilibrium statistical mechanics, see e.g. \cite{poincare}.
One can also decompose, in terms of its time symmetric components $t$ and its time antisymmetric components:
\[
\mathcal{A} = (\mathcal{T} - S)/2 \quad,
\]
where
\be
 S(\omega) = \mathcal{A}(g\omega) - \mathcal{A}(\omega) \hskip 5pt, \quad \mathcal{T}(\omega) = \mathcal{A}(g\omega) + \mathcal{A}(\omega) \label{revers} \quad .
\ee

and $g$ is the time reversal operator:
\be
 g \omega = ((\pi x)_{t-s}, 0\leq s\leq t) \quad , \label{invstoc}
\ee
with $\pi{x}$ equal to $x$ except for flipping any other variable with negative
parity under time reversal.
The quantity $S(\omega)$, under the assumption of local detailed balance \cite{Katz}, is the entropy flux triggered
by the perturbation and released into the environment \cite{CMW}. On the other hand, the quantity $\mathcal{T}(\omega)$
is referred to, in the literature, as \textit{dynamical activity} \cite{Maes1,MNW}, as it measures the reactivity and
instability of a trajectory. Dynamical activity is thus much more concerned with kinetics than with thermodynamics but
 it allows us to explore response around equilibrium beyond the linear regime. This shows also that the noise along in- and outgoing trajectories
 is crucial for the determination of state plausibilities \cite{land1,land2,land}.

A simple calculation yields the following general expression for the action pertaining to the process described by Eq. (\ref{diff}):
\be
\mathcal{A}(\omega)=\frac{\beta}{2}\int_0^t ds \left[F^p_s \cdot \chi F + \nabla \cdot (D F^p_s)+
\frac{1}{2} F^p_s \cdot \chi F^p_s\right] - \frac{\beta}{2}\int_0^t \id x_s \circ F^p_s
\label{A}
\ee
where the stochastic integral with the $\circ$ is in the sense of
Stratonovich. From (\ref{revers}) and (\ref{A}), one can derive the following
expressions for $S(\omega)$ and $\mathcal{T}(\omega)$:
\be
S(\omega) = \beta \int_0^t \id x_s \circ F^p_s \hskip 17pt \mbox{and} \hskip 17pt \mathcal{T}(\omega) = \mathcal{T}_1+\mathcal{T}_2  \nonumber \quad ,
\ee
with
\be
\mathcal{T}_1 =  \beta \int_0^t ds \left[F^p_s \cdot \chi F + \nabla \cdot (D
F^p_s)\right] \hskip 17pt \mbox{and} \hskip 17pt \mathcal{T}_2 =  \frac{\beta}{2} \int_0^t ds  F^p_s \cdot \chi F^p_s \nonumber \quad .
\ee
If the chosen observable $\phi$ is endowed with an even kinematical parity, the following linear response formula can be thus established \cite{Maes1}:
\bea
\lan \phi \ran_t-\lan \phi \ran_0 &\simeq & \lan \phi(x_t) S(\omega)\ran_0=-\lan \phi(x_0) S(\omega)\ran_0 =\nonumber\\
&=&-\int dx_0 f_0(x_0) \phi(x_0) \lan S(\omega) \ran^{x_0}_{0}\label{1LG} \quad .
\eea
The expression (\ref{1LG}) looks similar to the response formula (\ref{genresp}) obtained for deterministic systems,
with the entropy flux $S(\omega)$ taking the role of the observable $\Omega$ defined in Eq. (\ref{omegat})
\footnote{This is not surprising and indeed it is common. The fact is that both derivation are very formal and general
and only the evolution operators and the observables must appear.}.
The quantity $\lan S \ran^{x_0}_{0}$, in Eq. (\ref{1LG}), denotes the conditional expectation of the entropy flux $S(\omega)$ over $[0,t]$ given that the path started from the state $x_0$.
Its instantaneous flux is defined as \cite{CMW,MN}:
\be
\lan S\ran^{x_0}_{0}= \beta \int_0^t \langle w(x_s) \rangle^{x_0}_{0} ds \label{w} \quad ,
\ee
where $w(x_s)$ corresponds to the instantaneous (time-antisymmetric, random)
work made by the perturbation $F^p_t$.

\subsection{Nonequilibrium steady states}

By setting $F_{nc}\neq 0$, in Eq. (\ref{diff}), one spoils the time-reversibility of the reference dynamics. Therefore, given enough time, the reference
dynamics settles on a nonequilibrium steady state described by an invariant density
$f_0$ (usually not known).
In the steady state, one can use the definition of the probability current given in Eq.
(\ref{fp}), to define the \textit{information potential} $\mathcal{I}_{f}$ \cite{Baies,Prost} as:
\be
\mathcal{I}_{f}=-\nabla (\log f_0)=(\beta/\chi) u-\beta F \quad,
\label{infpot}
\ee
where $u\equiv j_{f}/f_0$ denotes a probability velocity.
From Eq. (\ref{infpot}), the large deviations method detailed in Ref.\cite{mattval} leads to the following general response function for nonequilibrium overdamped diffusion processes:
\be
R(t-s)=\chi\lan \left[-\nabla \cdot F^p_s(x_s)+
\mathcal{I}_{f}(x_s)\cdot F^p_s(x_s)\right] \phi(x_t)\ran_0 \quad .
\ee
In particular, if the perturbation takes the (time-independent) gradient form $F^p=\nabla V$, an easy calculation yields:
\be
R(t-s)=\beta\lan \left(u(x_s)\cdot\nabla V(x_s)\right)\phi(x_t)\ran_0-\beta \lan LV(x_s) \phi(x_t)\ran_0 \quad , \label{LV}
\ee
with $L=\chi F\cdot \nabla+\chi/\beta\nabla^2$.
Next, by using the adjoint generator \footnote{$L^*$ is defined with the help of the stationary distribution $f_0$: for any two state functions $a$ and $b$, $L^*$ is such that $\int dx f_0(x)a(x)L^* b(x)=\int dx f_0(x)b(x)L a(x)$. For detailed balance dynamics, in particular, one has $L^*=\pi L \pi$, where $\pi$ flips the variables which are odd under time reversal.} $L^*=L-2u\cdot \nabla$, one can suitably cast Eq. (\ref{LV}) into the equivalent form \cite{mattval}:
\be
R(t-s)=-\beta\lan \left(u(x_s)\cdot\nabla V(x_s)\right)\phi(x_t)\ran_0+\beta \frac{d}{ds}\lan \phi(x_t) V(x_s)\ran_0 \quad . \label{LV2}
\ee
It is worth remarking that the function $u(x)$, in (\ref{infpot}), is unknown in general. Nevertheless, Eq.
(\ref{LV}) is relevant at a formal level, because it shows that the response function
can be expressed in terms of a suitable correlation function computed wrt
reference stationary density characterizing the nonequilibrium steady state.\\
One also readily notices that Eq. (\ref{LV2}) produces the classical Kubo
formula (\ref{rham}) for $F_{nc} = 0$ (i.e. $u=0$) or when
describing the response in a reference frame moving with drift velocity $u$.

\section{Concluding remarks}
\label{concl}

We have summarised some of the main results of the theory of nonequilibrium systems.
We emphasized the physical questions and mechanisms lying behind the formalism presenting the various results in their
historical order. Research has, in fact, gradually moved from the analysis of equilibrium systems to dissipative ones, from the regime of small fluctuations to
large deviations. Along this challenging route, we also stressed similarity and difference between the different mathematical frameworks. In particular we noted the
reassuring fact that (microscopic) deterministic dynamics, discussed in Sec. \ref{sec5}, give rise to similar linear
response formulae as those of the (mesoscopic) stochastic dynamics, reviewed in Sec. \ref{sec6}. The resulting
thermodynamic behavior of the observable under consideration is indeed expected not to depend on the mathematical
framework used in the modelling, as long as the different frameworks describe the same phenomena.\\
Although a comprehensive understanding of the physics of nonequilibrium systems is still missing, we thus believe that a unifying framework is gradually emerging.


\end{document}